\documentclass[10pt,twocolumn,twoside]{IEEEtran}
\IEEEoverridecommandlockouts
\usepackage{cite}
\usepackage{algorithmic}
\usepackage{textcomp}
\usepackage{xcolor, soul}
\usepackage{graphicx,color}
\usepackage{amsfonts,amsthm,amssymb}
\usepackage{mathrsfs,amsmath,arydshln,latexsym}
\usepackage{cite,url}
\usepackage[bookmarks,colorlinks]{hyperref}
\usepackage[linesnumbered,ruled,vlined]{algorithm2e}
\usepackage{multirow,array,makecell}
\usepackage{stfloats}
\usepackage{balance}
\usepackage{fancyhdr}
\usepackage{bm}
\usepackage{enumitem}
\usepackage{pifont}
\usepackage{subcaption}
\usepackage{nomencl}
\usepackage{colortbl}
\usepackage{diagbox}
\usepackage{balance}
\usepackage{threeparttablex}
\usepackage{pifont}
\usepackage{physics}

\sethlcolor{yellow}
\setstcolor{green}
\setulcolor{red}

\allowdisplaybreaks[4]

\newcommand{\tabincell}[2]{\begin{tabular}{@{}#1@{}}#2\end{tabular}}

\begin{document}

	\title{Rydberg Atomic Quantum Receivers for Classical Wireless Communication and Sensing
	}

	\author{Tierui Gong,~\IEEEmembership{Member,~IEEE}, 
        Aveek Chandra,
		Chau Yuen,~\IEEEmembership{Fellow,~IEEE}, \\
        Yong Liang Guan,~\IEEEmembership{Senior Member,~IEEE},
        Rainer Dumke,
        Chong Meng Samson See,~\IEEEmembership{Member,~IEEE},\\
        Mérouane Debbah,~\IEEEmembership{Fellow,~IEEE},
        Lajos Hanzo,~\IEEEmembership{Life Fellow,~IEEE}
		\vspace{-1.2cm}
		\thanks{T. Gong, C. Yuen and Y. Guan are with School of Electrical and Electronics Engineering, Nanyang Technological University, Singapore 639798 (e-mail: trgTerry1113@gmail.com, chau.yuen@ntu.edu.sg, eylguan@ntu.edu.sg). A. Chandra and R. Dumke are with School of Physical and Mathematical Sciences, Nanyang Technological University, 637371, Singapore (e-mails: cqtavee@nus.edu.sg, rdumke@ntu.edu.sg). C. M. S. See is with DSO National Laboratories, Singapore 118225 (e-mail: schongme@dso.org.sg). M. Debbah is with the Center for 6G Technology, Khalifa University of Science and Technology, Abu Dhabi, United Arab Emirates (e-mail: merouane.debbah@ku.ac.ae). L. Hanzo is with School of Electronics and Computer Science, University of Southampton, SO17 1BJ Southampton, U.K. (e-mail: lh@ecs.soton.ac.uk).}
	}
	
	\maketitle

	\begin{abstract}
        Rydberg atomic quantum receivers (RAQRs) are emerging quantum precision sensing platforms designed for receiving radio frequency (RF) signals. It relies on creation of Rydberg atoms from normal atoms by exciting one or more electrons to a very high energy level, thereby making the atom sensitive to RF signals. RAQRs realize RF-to-optical conversions based on atom-light interactions relying on the so called electromagnetically induced transparency (EIT) and Autler–Townes splitting (ATS), so that the desired RF signal can be read out optically. The large dipole moments of Rydberg atoms associated with rich choices of Rydberg states facilitate an ultra-high sensitivity ($\sim$ nV/cm/$\sqrt{\text{Hz}}$) and an ultra-broadband tunability (direct-current to Terahertz). RAQRs also exhibit compelling scalability and lend themselves to the construction of innovative, compact receivers. Initial experimental studies have demonstrated their capabilities in classical wireless communications and sensing. To fully harness their potential in a wide variety of applications, we commence by outlining the underlying fundamentals of Rydberg atoms, followed by the principles and schemes of RAQRs. Then, we overview the state-of-the-art studies from both physics and communication societies. Furthermore, we conceive Rydberg atomic quantum single-input single-output (RAQ-SISO) and multiple-input multiple-output (RAQ-MIMO) schemes for facilitating the integration of RAQRs with classical wireless systems. Finally, we conclude with a set of potent research directions.
        \vspace{-1em}
	\end{abstract}

	\section{Introduction}
	
    Next-generation wireless systems are expected to support a variety of functionalities and applications. Integrated communication and sensing functionalities of next-generation wireless require particularly high sensitivity and data rates, while supporting large-scale connectivity and ultra-low latency. Therefore, the radio frequency (RF) receivers should possess broadband tunability and wideband processing capability. Based upon advanced antennas and integrated circuit (IC), cutting-edge RF receivers support signal receptions by well-calibrated antennas, filters, amplifiers, and mixers \cite{Moghaddasi2020Multifunction}. The sensitivity is limited by the receiver's bandwidth, noise figure, and the minimum signal-to-noise ratio (SNR) required to demodulate the desired signal. Given a specific sensitivity, strong RF reception capability can be attained by using multi-antenna based multiple-input multiple-output (MIMO) schemes and sophisticated signal processing methods to combat channel fading and noise contamination. To receive wideband RF signals at different frequencies, both broadband tunability and wideband processing capability may be realized by stacking several IC branches along with extra metallic components, where each branch is responsible for a specific frequency band. Such RF systems are however band-limited and bulky, power-thirsty, and suffer from numerous design challenges, especially for large-scale, high-frequency receivers.

    The Rydberg atomic quantum receiver (RAQR) concept emerges as a radical solution to these RF reception challenges \cite{schlossberger2024rydberg,zhang2024rydberg,yuan2023quantum,Fancher2021Rydberg}. It relies on the creation of Rydberg states (excitation), which have remarkably high dipole moment and hence are eminently suitable for sensing the impinging RF signals. The response to an RF signal in RAQR is detected optically via change in the atomic spectroscopy signal. The RF field's amplitude measurements are directly linked to the International System of Units (SI) and they set absolute (atomic) standards for the RF electric fields from a metrology standpoint. Additionally, they are capable of directly down-converting RF signals, spanning from direct-current (DC) to Terahertz (THz), to the baseband without using any mixer, hence significantly simplifying the receiver’s structure and yielding a compact form-factor relative to conventional counterparts.

	To fully unlock the potential of RAQRs in communication and sensing, we offer a panoramic overview of RAQRs. We commence by highlighting the fundamentals of Rydberg atoms and proceed to powerful receivers, including their principles, schemes, impairments, and their critical comparisons to antenna-based receivers. We then summarize their capabilities and distinctive features, surveying typical studies from diverse research communities. Furthermore, we conceive the Rydberg atomic quantum single-input single-output (RAQ-SISO) and RAQ-MIMO schemes to facilitate the integration of RAQRs with existing classical wireless systems. We finally forecast several promising future directions in this very first article on RAQRs in wireless communication and sensing from an engineering perspective.

    \begin{figure*}[t!]
    	\centering
    	\includegraphics[width=0.89\textwidth]{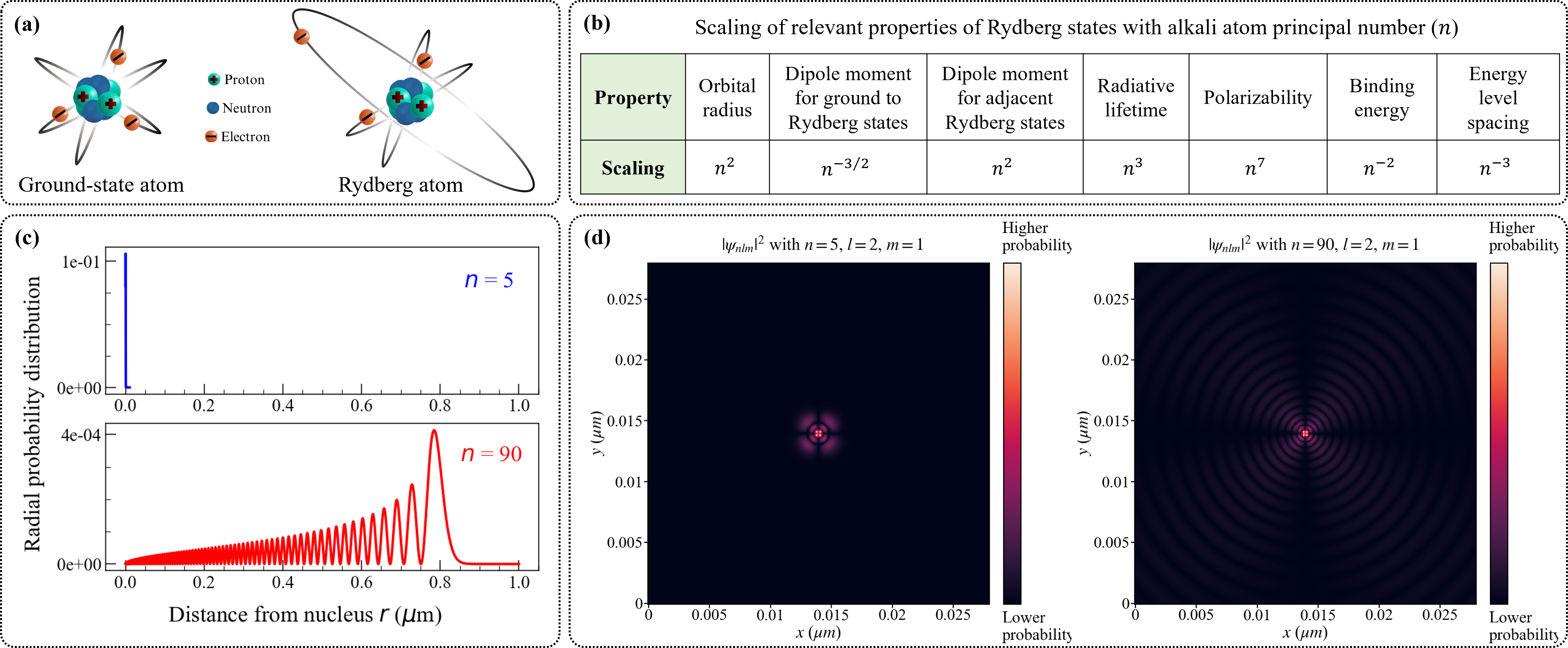}
    	\caption{(a) A simplified diagram of atomic structure for ground-state and Rydberg atoms. (b) Scaling of the relevant properties of Rydberg states with the alkali atom principal number ($n$). (c) Radial probability distribution of Rydberg states of Cs atoms for $n=5, 90$ ($l = 2$). (d) The electron probability distribution $|\psi_{nlm}|^2$ for an atom excited to a state $n=5$, $l=2$, $m=1$ and for a Rydberg atom in state $n=90$, $l=2$, $m=1$, respectively.}
    	\label{fig:Rydberg}
    	\vspace{-1em}
    \end{figure*}

    \section{Rydberg Atomic Quantum Receiver}
    \label{Section_CCM}
    RAQRs exploit the properties of Rydberg atoms, hence we commence by their brief overview, followed by their principles and schemes. 
    Finally, we present the impairments of RAQRs, and compare them to classical RF receivers.

	\subsection{Fundamentals of Rydberg Atoms}
	\subsubsection{Atomic Structure \cite{Gallagher94}}
    Quantum mechanics provides a comprehensive description of the structure, properties and behavior of atoms. Each atom has a nucleus at its core and there are electrons that orbit around the nucleus. The energy of electrons in these orbits is quantized, i.e., they occupy only certain discrete energy-levels and cannot be traced in between these levels. However, electrons can transition from one energy level to another by absorbing or emitting a photon with energy equal to the difference between the energy levels. Electrons have a wave-particle dual nature. In the quantum mechanical description, the state of an electron is described by a wavefunction $\psi$, which contains all information about the electron's position, energy, spin, orientation, etc. The square of the wavefunction, $|\psi|^2$ gives the probability density (i.e. probability per unit volume) of finding the electron in a certain spatial region around the nucleus. These regions are termed as orbitals.
    The unique quantum state of an electron is described by four quantum numbers:
       \begin{itemize}
        \item \textbf{Principal quantum number $\bm{n}$} (positive integer $n$): It indicates the energy level or shell which the electron occupies. A higher $n$ implies that the electron resides further away from the nucleus. 
        
        \item \textbf{Azimuthal or orbital angular momentum quantum number $\bm{l}$} (non-negative integer $l\le n-1$): It accounts for the sub-shell (sub-level) that the electron occupies, with each sub-shell having unique characteristic shapes. 
        
        \item \textbf{Magnetic quantum number $\bm{m}$} (integer $-l \leq m \leq l$): It indicates the specific orbital that an electron occupies, which corresponds to a certain spatial orientation.
        
        \item \textbf{Spin quantum number $\bm{s}$} ($s = \pm \frac{1}{2}$): It denotes the orientation of an electron's intrinsic spin. Following Pauli's exclusion principle, two electrons in an identical orbital must have opposite spin orientations.
        \end{itemize}

    If mediated by photons of electromagnetic (EM) field, the transition of an electron from one quantum state to another is called an electric dipole transition. Based on the conservation of angular momentum and parity\footnote{Parity refers to the symmetry of a wavefunction under spatial inversion.}, only those (electric dipole) transitions are allowed which satisfy the selection rules: \ding{172} $\Delta l = \pm1$, \ding{173} $\Delta m = 0, \pm 1$, and \ding{174} the initial and final states must have opposite parities.

	\subsubsection{Properties of Rydberg State}
	Alkali atoms have single valence electrons in their ground state in the outermost orbital. This represents a relatively loose bound and therefore it can be excited to a high energy level associated with a high principal quantum number, of say $n>10$ for atomic systems. When such an excitation is created, the atom and its corresponding state are Rydberg ones. The atomic structures of the ground-state and Rydberg-state atoms are schematically shown in Fig. \ref{fig:Rydberg}(a). 
	The Rydberg states of any alkali atom and a hydrogen atom have essentially similar structures and properties. In both cases, a single electron occupies a high-$n$ state and has an effective core of unit positive charge at the center. The only difference lies in the finite size of the positively charged core, which is larger for alkali atoms relative to hydrogen atoms. This leads to a small correction (via quantum the defect parameter $\delta_l$) to the energies for non-hydrogenic atoms in particular for low $l$ states. But again, one can say that the general properties of all Rydberg atoms are quite similar. The scaling of the relevant properties of Rydberg states with $n$ is shown in Fig. \ref{fig:Rydberg}(b). The $n$-dependence arises from the Rydberg atom wavefunction characteristics. For RAQRs, the typical choice of alkali atoms falls on either Caesium (Cs) or Rubidium (Rb) due to the availability of commercial laser systems for the transition wavelengths of these species.

	The Rydberg atoms are large, since the average distance from the nucleus $\langle r\rangle$ scales with $n^2$, resulting in large dipole matrix elements. This point is illustrated in Fig. \ref{fig:Rydberg}(c), where the radial probability distribution $|rR_{nl}(r)|^2$ for state $nS$ of Cs atoms is plotted against $r$ (distance from nucleus) for increasing $n$. The atomic size increases dramatically with $n$ and its value approaches 1 $\mu m$ for $n \ge 100$, while it is about 0.1 nm for the ground-state atom. Since Rydberg atoms are hydrogen-like, the analytical solution of Schr$\ddot{o}$dinger's equation for the hydrogen atom can help us to visualize the Rydberg atom wavefunctions in space. In Fig. \ref{fig:Rydberg}(d), the spatial extent and characteristic shape of the electron probability distribution $|\psi_{nlm}|^2$ for an atom excited to a state $n=5$, $l=2$, $m=1$, has been compared to a Rydberg atom in state $n=90$, $l=2$, $m=1$. The non-zero probability of finding the electron at few hundred nanometers away is evident for a Rydberg atom, while this probability is zero for a normal excited-state atom.

    \begin{figure}[t!]
		\centering
		\includegraphics[width=0.486\textwidth]{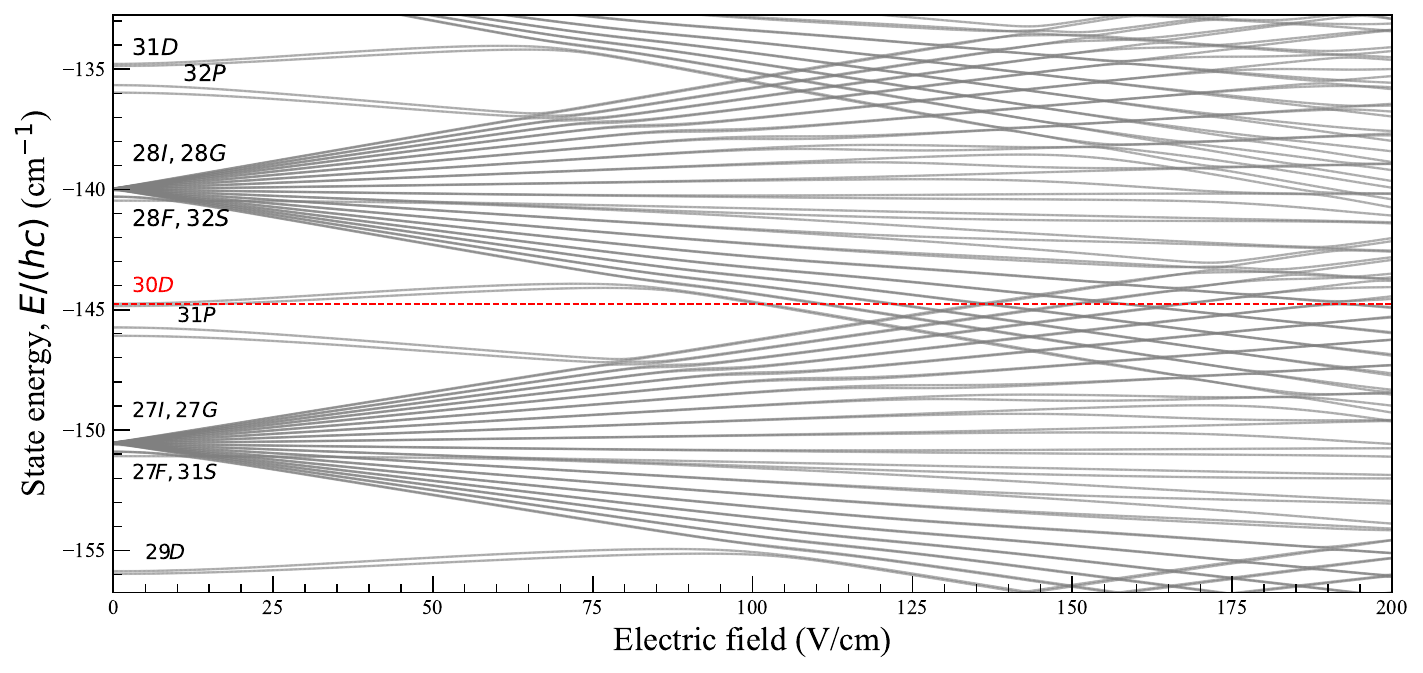}
		\caption{Stark map of Cs atoms, in the vicinity of the $30D$ state, showing the energy shifts experienced by different states in the presence of electric field. The energy level $30D$ in the absence of electric field is marked by red dashed line}
		\label{fig:starkmap}
		\vspace{-1em}
    \end{figure}

	\subsubsection{Rydberg Atoms in Electric Fields}
	The sensitivity of Rydberg atoms to electric fields arises due to the high polarizability of Rydberg states. In the presence of electric field, the energy levels shift for different states in a way similar to the energy-level splitting of a hydrogen atom. This is called the Stark effect and the associated Stark shifts can be calculated by direct diagonalization of the Hamiltonian matrix. The plot of atomic energy levels as a function of the electric field strength applied is termed as the Stark map and such a map is shown in Fig. \ref{fig:starkmap} for Cs atoms in the vicinity of the $30D$ state. 
	The red dashed line shows the energy level in the absence of electric field. The low-$l$ ($l=0,1,2$) Rydberg states are non-degenerate and experience a second-order Stark shift, i.e., a quadratic scaling with electric field amplitude $\mathcal{E}$ given by $\Delta E_{\text{DC}} = \alpha \mathcal{E}^2/2$, where $\alpha$ is the DC polarizability of the Rydberg state. For $l>3$, the states are degenerate and therefore experience a linear Stark shift similar to the hydrogen Stark spectrum.

	When the electric field interacting with atoms is constant or fluctuates at a low frequency, it is called DC Stark shift, as mentioned above. When a high-frequency electric field interacts with the atoms, it results in AC Stark shift, which has quadratic scaling with the field amplitude $\mathcal{E}$. The magnitude and sign of the AC Stark shift depends on the detuning $\Delta$ of the electric field from the resonant transition frequency and scales as $\Delta E_{\text{AC}}\propto 1/\Delta$. This implies that the magnitude of the shift is larger when the electric field is closer to resonance, while its sign is positive (negative) for blue-detuned (red-detuned) electric fields. The Stark effect causes coupling between closely-spaced energy levels leading to the mixing of energy eigenstates. 
	As a result of mixing the of wavefunctions of two states anti-crossings appear in the Stark map, as a gap between two energy curves where they would otherwise cross each other. This has been explained in detail in \cite{Gallagher94}.

	\subsection{Principles and Schemes of RAQRs}
	\label{Sec:Principal&Scheme}
	
	\subsubsection{Rydberg-Electromagnetically Induced Transparency (EIT) Spectroscopy and the Standard Scheme\cite{Fancher2021Rydberg}}
	\label{sec:standard}
	EIT is a quantum interference phenomenon, where two excitation pathways of a three-level system interfere to open a transmission window in the absorption spectrum of atomic resonance. This is typically fulfilled by counter-propagating a pair of laser beams, each of which corresponds to an excitation pathway. Since this is a coherent process, it is ultra-sensitive to phase perturbations and energy level shifts of the three-level system.

	Rydberg excitation is prepared via a two-photon excitation process, where a pair of counter-propagating laser beams of different wavelengths, namely the `probe' and `coupling (control)', are overlapped inside an alkali vapor cell, as shown in Fig. \ref{fig:AtomSensor}(a). This is accomplished via Rydberg-EIT in a ladder-type configuration, namely $\ket{1}$$\rightarrow$$\ket{2}$$\rightarrow$$\ket{3}$$\rightarrow$$\ket{4}$ (`standard') in Fig. \ref{fig:AtomSensor}(h). The probe beam excites the atom from its ground state to an intermediate excited state, while the coupling beam takes it to a high-energy (principal number $n$) state, namely the Rydberg state. When an external RF signal's frequency is (near) resonant with two adjacent Rydberg states, changes can be observed in the optical spectroscopic signal in the following three scenarios, namely (i) probe beam only, (ii) probe \& coupling beams, and (iii) probe \& coupling beams plus RF signal, as illustrated in Fig. \ref{fig:AtomSensor}(b) using realistic experimental parameters. With a further increase in the RF field strength, the Rydberg-EIT peak splits into two, known as the Autler-Townes splitting (ATS). The RF signal strength can be determined by measuring the frequency splitting of the two peaks of ATS.

	This is the standard scheme (structure) used in RAQRs. The electron transitions take place among the four energy levels (quantum states) $\ket{1}$, $\ket{2}$, $\ket{3}$, and $\ket{4}$ with resonant transition frequencies $\omega_{(P,C,s)}$ and detunings $\Delta_{(P,C)}$, as shown in Fig. \ref{fig:AtomSensor}(h). Each quantum state has its own decay rate, denoted by $\gamma$. The transition $\ket{1}\rightarrow \ket{2}$ is typically chosen to be D2 transition of Cs (Rb). $\ket{3}$ could be any Rydberg state, and the excitation is created by tuning the coupling laser wavelength to the appropriate value. Finally, $\ket{4}$ in the `standard' box of Fig. \ref{fig:AtomSensor}(h) is the neighboring Rydberg state and its choice is governed by the particular RF frequency that the RAQR would be set up to receive. In this scheme, the measurement sensitivity approaches $\sim$ \textmu V/cm/$\sqrt{\text{Hz}}$ \cite{schlossberger2024rydberg}.

	\subsubsection{Superheterodyne Scheme for Enhanced Sensitivity and Phase Extraction \cite{holloway2019detecting,jing2020atomic}}
	\label{sec:superheterodyne}
	RAQR's measurement sensitivity can be further improved by exploiting the superheterodyne principle, where a local oscillator (LO) as a reference is added to the structure for detecting relatively weak RF signals. This scheme is presented in Fig. \ref{fig:AtomSensor}(c). The RF signal to be detected can be at intermediate frequency within the instantaneous bandwidth of RAQRs. 
	The LO facilitates the creating of microwave-dressed Rydberg states and the frequency splitting measurement applied in the standard scheme is converted into an amplitude-modulated measurement in the presence of these two RF signals. As such, Rydberg atoms serve as the role of a conventional RF mixer. In this scheme, the phase of the desired RF signal can be measured relative to that of the LO from the modulated amplitude of the probe beam. The sensitivity has been improved to 55 nV/cm/$\sqrt{\text{Hz}}$ \cite{jing2020atomic}.

	\subsubsection{Optical Laser Array (OLA) for Enhanced Sensitivity \cite{wu2024enhancing}}
	\label{sec:OLA}
	The sensitivity of the superheterodyne scheme can be further enhanced by exploiting the OLA structure (Fig. \ref{fig:AtomSensor}(d)), where multiple laser beams are generated in parallel by employing well-designed gratings (split-beam and collimator gratings). Therefore, multiple parallel Rydberg atomic regions are created for simultaneously receiving RF signals. The multiple laser beams are combined and detected by the photodetector (PD) for realizing high-sensitivity detectors. More particularly, the sensitivity has been enhanced to 19 nV/cm/$\sqrt{\text{Hz}}$ \cite{wu2024enhancing}. By employing the OLA scheme to the most advanced Rydberg sensor, the sensitivity of $<$ 1 nV/cm/$\sqrt{\text{Hz}}$ is achievable \cite{wu2024enhancing}.

	\subsubsection{Polarization-Sensitive Scheme (PSS) for Vector Field Sensing \cite{elgee2024complete}}
	\label{sec:polarization}
\begin{figure*}[t!]
		\centering
		\includegraphics[width=0.88\textwidth]{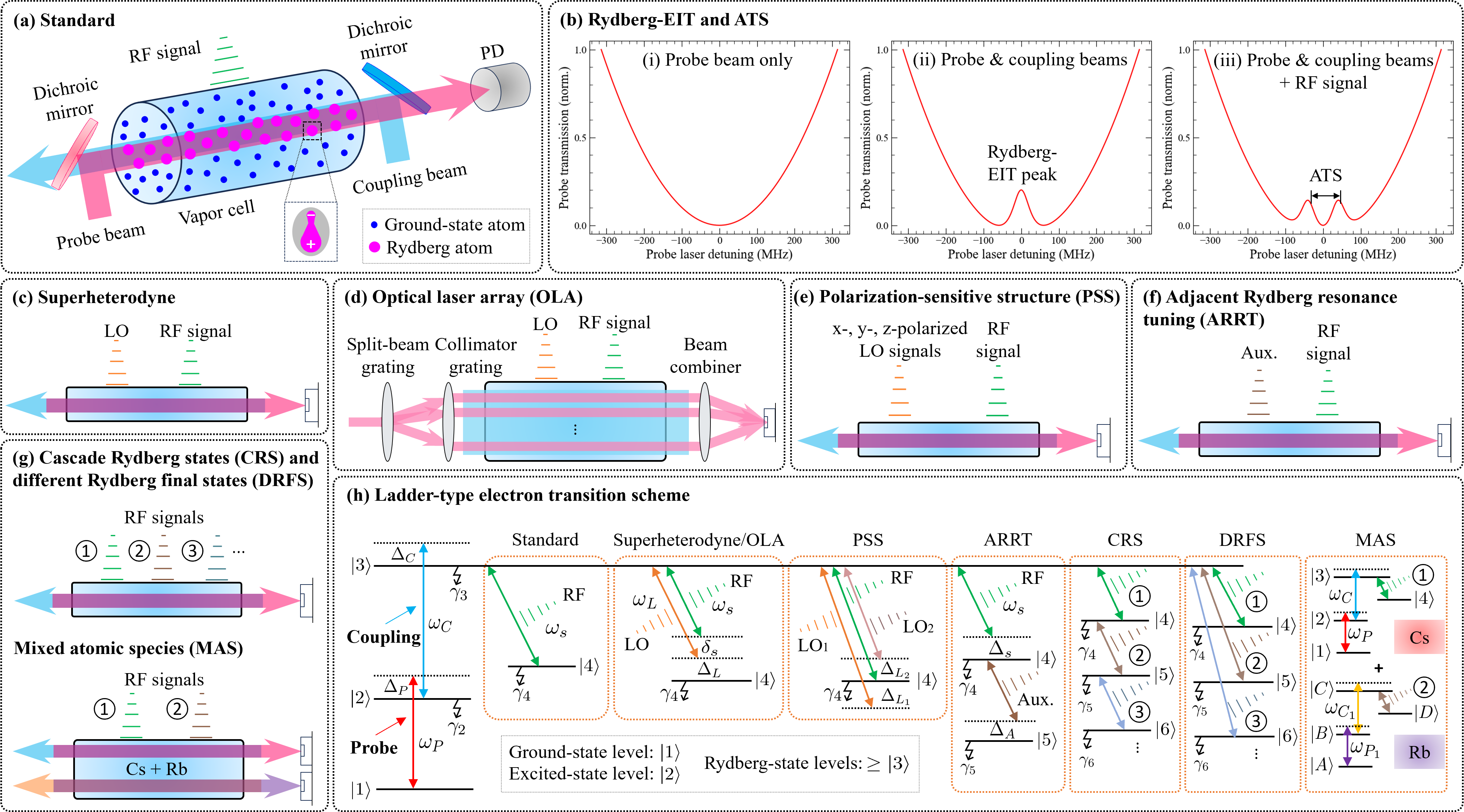}
		\caption{(a) The standard scheme, and (b) Rydberg-EIT spectroscopy signal and ATS. (c) The superheterodyne scheme, (d) OLA-based scheme, (e) PSS scheme, (f) continuous-band detection scheme, and (g) multiband detection schemes. (h) The ladder-type scheme for showing the electron transition of the above schemes.}
		\label{fig:AtomSensor}
		\vspace{-1em}
	\end{figure*}

	The polarization of RF signals can be determined by RAQRs by exploiting three independent and orthogonally-polarized LOs. Specifically, these LOs are $x$-, $y$-, and $z$-polarized, respectively, as seen in Fig. \ref{fig:AtomSensor}(e). The ladder-type scheme is presented in the `PSS' box of Fig. \ref{fig:AtomSensor}(h). The RF signal to be detected is mixed with these LO signals and thus three signal beats are created for extracting their relative phases. Notably, these relative phases facilitate the determination of the signal polarization.

	\subsubsection{Adjacent Rydberg Resonance Tuning (ARRT) for Enhanced Instantaneous Bandwidth and Continuous-Band Detection \cite{simons2021continuous,liu2022continuous}}
	\label{sec:CBD}
	The RF signal to be detected is typically limited to a narrow instantaneous bandwidth ($\le$ 10 MHz \cite{Fancher2021Rydberg}) around the transition frequency of coupled Rydberg states. The reason for this is that a wide bandwidth leads to substantial RF detuning, creating asymmetric ATS peaks. Once the RF detuning exceeds some narrow range, the RF signal is no longer resonant with the Rydberg state and the ATS cannot be observed. Extending the instantaneous bandwidth is feasible by exploiting the ARRT scheme of \cite{simons2021continuous}, where an auxiliary RF signal is employed to couple another Rydberg state adjacent to that coupled with the desired RF signal, as shown in Fig. \ref{fig:AtomSensor}(f) and the `ARRT' box of Fig. \ref{fig:AtomSensor}(h). The instantaneous bandwidth can be up to $100$ MHz \cite{zhang2024rydberg} and reference [39] of \cite{zhang2024rydberg}. This scheme facilitates the detection of RF signals distributed across a continuous frequency range (e.g., $\ge 1$ GHz \cite{liu2022continuous}).

	\subsubsection{Cascade Rydberg States (CRS) \cite{allinson2024simultaneous}, Different Rydberg Final States (DRFS) \cite{meyer2023simultaneous}, and Mixed Atomic Species (MAS) \cite{holloway2021multiple} for Simultaneous Multiband Detection}
	\label{sec:MDBD}
	RAQRs are capable of simultaneously detecting multiple RF signals having different carrier frequencies by exploiting three different schemes, as shown in Fig. \ref{fig:AtomSensor}(g) and the `CRS', `DRFS', `MAS' boxes of Fig. \ref{fig:AtomSensor}(h). Specifically, the CRS scheme \cite{allinson2024simultaneous} exploits a series of high-orbital-angular-momentum states that are simultaneously cascaded by multiple RF signals having their carrier frequencies in descending order (from THz to MHz). Furthermore, the DRFS scheme \cite{meyer2023simultaneous} simultaneously harnesses multiple transitions from an identical Rydberg state to multiple different Rydberg states, where each transition corresponds to the RF signal having a specific frequency. Lastly, the MAS scheme \cite{holloway2021multiple} realizes the simultaneous detection by exploiting mixed alkali atomic species (Cs and Rb). Each atom species is excited relying on a pair of laser beams having specific wavelengths and hence detecting the corresponding RF signal having a specific frequency.

	\subsubsection{Other Promising Schemes}
	Beyond the above schemes, a suite of more sophisticated schemes emerge. 
	For instance, by exploiting the machine learning models \cite{liu2022deepNC}, the RAQRs can be trained to better receive RF signals than those schemes relying on the master equation. Additionally, the scheme exploiting the many-body effect exhibits an enhanced sensitivity, as seen in \cite{zhang2024rydberg} and reference [124] therein, paving the way for approaching the standard quantum limit (SQL). Furthermore, other advanced schemes exploiting Schrodinger cat states and squeezed states may potentially allow RAQRs to surpass the SQL, as discussed in \cite{yuan2023quantum} and references [101], [103] therein.


	\subsection{Sources of Impairments}
	\label{sec:impairments}
	The performance of RAQRs in communication and sensing is influenced by several impairments, as detailed below.

	\subsubsection{RAQR Noise Sources}
	Several factors contribute to the noise of RAQRs. Among them, the photon shot noise and the electrical noise from the photodetector (PD) may increase the noise floor. The laser noise plays a crucial role in determining the sensitivity of RAQR. The frequency noise of lasers can be minimized by frequency stabilization techniques, where the laser is locked to the mode of a high-finesse ultra-low expansion cavity. The intensity noise of lasers can be reduced similarly by implementing a servo feedback loop and a balanced photodetection scheme. The vibrational instability of mirrors may contribute to the mechanical noise. All these can be mitigated by innovative engineering techniques.

	\subsubsection{Transit Time and Doppler Broadening \cite{schlossberger2024rydberg}}
	The atoms in the vapor cell only interact with laser beams during the so called transit time. They will leave the interaction region due to their thermal motion (kinetic energy) at room temperature. This leads to the broadening of spectral lines termed as transit time broadening. An atom moving in (against) the direction of light beam  will experience an energy level shift i.e., redshifted (blueshifted) with respect to the light frequency. When this frequency is resonant to an atomic transition, light gets absorbed by the atoms obeying the same (Gaussian) velocity distribution of the atoms. These different velocities result in the broadening of spectral lines, termed as Doppler broadening. In RAQRs, there is a mismatch in wavelengths for the counter-propagating probe and coupling laser beams which accounts for the Doppler-broadened Rydberg-EIT peak. The broadening is proportional to the residual Doppler mismatch. 

	\subsubsection{Collisional Effects \cite{schlossberger2024rydberg}}
	The interaction or collision of Rydberg atoms with the vapor-cell wall becomes significant at roughly micron-scale distances, since there are boundary conditions imposed on the atomic dipoles, as they approach the wall. There are other collisions in this system, like Rydberg atom - Rydberg atom collisions and ground-state - Rydberg atom collisions, that play an important role. All of these collisions cause dephasing of the quantum states and reduce the coherence time $T_r$ of the RAQR receiver. It has been shown in \cite{schlossberger2024rydberg} that when all these dephasing effects are taken into account, the coherence time $T_r$ and hence the sensitivity reaches the maximum value for Rydberg states in the range of $n=45, \cdots, 60$.

	\subsubsection{Quantum Projection Noise \cite{Fancher2021Rydberg}}
	It is also known as the quantum shot noise and often referred to as SQL, since it sets a lower limit to RAQRs' sensitivity. This noise arises from the quantum fluctuations of the measurement process. The SQL scales as $\propto 1/\sqrt{NT_r T_i}$, where $N$ is the number of uncorrelated Rydberg atoms participating in RF reception, $T_i$ is the integration time and $T_r$ is the coherence time of the RAQR. The typical SQL sensitivity is on the order of pV/cm/$\sqrt{\text{Hz}}$ \cite{jing2020atomic}.

	\subsubsection{Noise from RF Background}
	\label{subsubsec_Noise_RF}
	RAQRs cannot differentiate between an RF target signal and the background electromagnetic noise. The change in the response of RAQR in the presence of bandwidth-limited white Gaussian RF noise has been theoretically modelled and verified experimentally, as seen from \cite{zhang2024rydberg} and reference [98] of \cite{zhang2024rydberg}.

	\subsection{Comparison to Conventional RF Receivers}

	\subsubsection{Operating Principle}
	Traditional RF receivers capture incident electric fields by using antennas and generate the corresponding currents. The size of antennas requires to match the wavelength of the detected electric field. By contrast, RAQRs realize RF reception via Rydberg atoms created by driving laser beams in a vapor cell. The size of RAQRs is independent to the wavelength.

	\subsubsection{SI-Traceability and Calibration}
	Conventional antennas need fine calibration for high-integrity reception. The calibrations are performed by placing uncalibrated antennas in a known EM environment that is measured via a well-calibrated antenna, leading to a `chicken-or-egg' dilemma. However, RAQRs allow precise measurements of RF signals which is directly related to Planck's constant and hence can self-calibrate by relying on SI-traceable atomic standards.

	\subsubsection{Passband-Baseband Conversion}
	Classical RF receivers, exploiting the popular homodyne/heterodyne/superheterodyne RF structures, require tightly-integrated RF front-end components (e.g., mixers, amplifiers, filters) for down-converting the passband RF signal to baseband, requiring complex integrated circuits, especially for large-scale systems \cite{Moghaddasi2020Multifunction}. By contrast, the measurement scheme of RAQRs allow direct down-conversion of RF signals \cite{holloway2021multiple}, bypassing the complex integrated circuitry, thus facilitating a simple, compact structure.

	\subsubsection{Receiver Impairments}
	Classical antenna-based receivers suffer from Gaussian thermal noise and electronic hardware impairments, while RAQRs suffer from several different impairments, ranging from the noise of lasers and PDs to noise arising from broadening and dephasing effects in atomic systems, as explained in Section \ref{sec:impairments}. 
	The sensitivity of RAQRs is set by the SQL \cite{schlossberger2024rydberg,zhang2024rydberg,yuan2023quantum,Fancher2021Rydberg}, whereas the sensitivity of their counterparts is determined by the thermal noise limit, as discussed in \cite{schlossberger2024rydberg} and reference [39] of \cite{schlossberger2024rydberg}.

	\subsubsection{Angular Reception Ranges} 
	Antenna (array-) based receivers can only receive impinging RF signals from a restricted spatial angle, e.g., half-plane range to the greatest extent for antenna arrays. By contrast, RAQRs are capable of receiving signals from wider spatial angles with lesser restrictions, facilitating the reduction of sensor deployments.

	\subsubsection{Mutual Coupling Effects}
	Conventional antenna arrays typically face negative mutual coupling effects, which become more serious as antenna arrays become dense and large. The mutual coupling effects may cause considerable challenges for receiver designs. Fortunately, RAQRs are less susceptible to mutual coupling as Rydberg sensors do not re-emit the impinging RF signal, facilitating simplified receiver designs.

   \section{RAQR Enabled Quantum-Aided Wireless Communication and Sensing}
	\label{Section_NFCM}
        The application of RAQRs to classical wireless communication and sensing is promising, as indicated by representative studies from diverse research communities that promote quantum-aided classical wireless communication and sensing.

	\begin{table*}[!t]
		\renewcommand{\arraystretch}{1.16}
		\caption{\textsc{State-of-the-art in RAQRs.}}
		\label{tab:State-of-the-art}
		\centering
		\tabcolsep = 0.1cm
		
		\resizebox{\linewidth}{!}{
			\begin{tabular}{!{\vrule width0.6pt}c|c|c|c|c|c|c|c|c|l!{\vrule width0.6pt}}			
				
				\Xhline{0.6pt}
				\rowcolor{yellow!60} 
				\multicolumn{10}{|c|}{\textbf{Theoretical and experimental studies from the physics community}} \\
				
				\Xhline{0.6pt}
				\rowcolor{cyan!15}
				\textbf{\tabincell{c}{Ref.}}
				& \textbf{\tabincell{c}{Structure}} 
				& \textbf{\tabincell{c}{Capability}} 
				& \textbf{\tabincell{c}{Features}} 
				& \textbf{\tabincell{c}{Atom}} 
				& \textbf{\tabincell{c}{Energy levels}} 
				& \multicolumn{2}{c|}{\cellcolor{cyan!15} \textbf{\tabincell{l}{Laser (probe/coupling)\\ Power \;\;\;\;\;\; Wavelength}}}
				& \textbf{\tabincell{c}{RF frequency}} 
				& \qquad \qquad \textbf{\tabincell{c}{Main Contributions}} \\

			\Xhline{0.6pt}
			\tabincell{c}{\cite{holloway2019detecting}} 
			& \tabincell{c}{Superheterodyne} 
			& \tabincell{c}{C1, C2, C4\\ (BPSK, QPSK, QAM)} 
			& \tabincell{c}{F1, F3} 
			& \tabincell{c}{\textsuperscript{\scalebox{0.8}{133}}Cs} 
			& \tabincell{c}{
			6S\textsubscript{\scalebox{0.8}{1/2}} \textrightarrow 6P\textsubscript{\scalebox{0.8}{3/2}} \textrightarrow \\
			34D\textsubscript{\scalebox{0.8}{5/2}} \textrightarrow 
			35P\textsubscript{\scalebox{0.8}{3/2}} } 
			& \tabincell{c}{41.2 \textmu W /\\ 48.7 mW} 
			& \tabincell{c}{850.53 nm /\\ 511.148 nm} 
			& \tabincell{c}{LO: 19.629 GHz,\\ IF: 500 kHz \& 1 MHz} 
			& \tabincell{l}{The BPSK, QPSK, and QAM modulated\\ signals are able to be received by RAQRs.}\\ 

				\Xhline{0.6pt}
				\tabincell{c}{\cite{jing2020atomic}} 
				& \tabincell{c}{Superheterodyne} 
				& \tabincell{c}{C1, C2} 
				& \tabincell{c}{F1} 
				& \tabincell{c}{\textsuperscript{\scalebox{0.8}{133}}Cs} 
				& \tabincell{c}{
					6S\textsubscript{\scalebox{0.8}{1/2}} \textrightarrow 6P\textsubscript{\scalebox{0.8}{3/2}} \textrightarrow \\
					47D\textsubscript{\scalebox{0.8}{5/2}} \textrightarrow 
					48P\textsubscript{\scalebox{0.8}{3/2}} } 
				& \tabincell{c}{120 \textmu W /\\ 34 mW} 
				& \tabincell{c}{852 nm /\\ 510 nm} 
				& \tabincell{c}{LO: 6.94 GHz,\\ IF: 150 kHz} 
				& \tabincell{l}{Experimentally illustrate a sensitivity of\\ 55 nV/cm/$\sqrt{\text{Hz}}$ by superheterodyne.}\\

				\Xhline{0.6pt}
				\tabincell{c}{\cite{wu2024enhancing}} 
				& \tabincell{c}{OLA,\\ superheterodyne} 
				& \tabincell{c}{C1, C2} 
				& \tabincell{c}{F1} 
				& \tabincell{c}{\textsuperscript{\scalebox{0.8}{133}}Cs} 
				& \tabincell{c}{
					6S\textsubscript{\scalebox{0.8}{1/2}} \textrightarrow 6P\textsubscript{\scalebox{0.8}{3/2}} \textrightarrow \\
					44D\textsubscript{\scalebox{0.8}{5/2}} \textrightarrow 
					45P\textsubscript{\scalebox{0.8}{3/2}} } 
				& \tabincell{c}{30 \textmu W /\\ 135 mW} 
				& \tabincell{c}{852 nm /\\ 509 nm} 
				& \tabincell{c}{LO: 8.57 GHz} 
				& \tabincell{l}{Experimentally illustrate a sensitivity of\\ 19 nV/cm/$\sqrt{\text{Hz}}$ by proposing the OLA.}\\

				\Xhline{0.6pt}
				\tabincell{c}{\cite{elgee2024complete}} 
				& \tabincell{c}{PAS} 
				& \tabincell{c}{C1, C2, C3} 
				& \tabincell{c}{F1, F6} 
				& \tabincell{c}{\textsuperscript{\scalebox{0.8}{85}}Rb} 
				& \tabincell{c}{
					5S\textsubscript{\scalebox{0.8}{1/2}} \textrightarrow 5P\textsubscript{\scalebox{0.8}{3/2}} \textrightarrow \\
					60D\textsubscript{\scalebox{0.8}{5/2}} \textrightarrow 
					59F\textsubscript{\scalebox{0.8}{7/2}} } 
				& \tabincell{c}{N/A} 
				& \tabincell{c}{780 nm /\\ 480 nm} 
				& \tabincell{c}{LO\textsubscript{x} 10.6633 GHz,\\
					LO\textsubscript{y} 10.66327 GHz,\\
					LO\textsubscript{z} 10.6628 GHz} 
				& \tabincell{l}{Experimentally illustrate a sensitivity of\\ 57 \textmu V/m/$\sqrt{\text{Hz}}$ and 66 \textmu V/m/$\sqrt{\text{Hz}}$ for\\ the horizontal and vertical polarizations.}\\

				\Xhline{0.6pt}
				\tabincell{c}{\cite{simons2021continuous}} 
				& \tabincell{c}{ARRT} 
				& \tabincell{c}{C1, C5\\ (continuous-band)} 
				& \tabincell{c}{F1, F2,\\ F4, F6} 
				& \tabincell{c}{\textsuperscript{\scalebox{0.8}{133}}Cs} 
				& \tabincell{c}{
					6S\textsubscript{\scalebox{0.8}{1/2}} \textrightarrow 6P\textsubscript{\scalebox{0.8}{3/2}} \textrightarrow \\
					68S\textsubscript{\scalebox{0.8}{1/2}} \textrightarrow 
					67P\textsubscript{\scalebox{0.8}{3/2}}  
					(68P\textsubscript{\scalebox{0.8}{3/2}}) \\ \textrightarrow 
					65D\textsubscript{\scalebox{0.8}{5/2}}  (66D\textsubscript{\scalebox{0.8}{5/2}}) } 
				& \tabincell{c}{N/A} 
				& \tabincell{c}{850 nm /\\ 509 nm} 
				& \tabincell{c}{11.25 GHz -\\ 13.9 GHz} 
				& \tabincell{l}{The ARRT scheme was demonstrated for\\ detecting continuous-band signals.}\\

				\Xhline{0.6pt}
				\tabincell{c}{\cite{liu2022continuous}} 
				& \tabincell{c}{ARRT,\\superheterodyne} 
				& \tabincell{c}{C1, C2, C4\\ (QPSK), C5\\ (continuous-band)} 
				& \tabincell{c}{F1, F2,\\ F4, F6} 
				& \tabincell{c}{\textsuperscript{\scalebox{0.8}{87}}Rb} 
				& \tabincell{c}{
					5S\textsubscript{\scalebox{0.8}{1/2}} \textrightarrow 5P\textsubscript{\scalebox{0.8}{3/2}} \textrightarrow \\
					70D\textsubscript{\scalebox{0.8}{5/2}} \textrightarrow 
					71P\textsubscript{\scalebox{0.8}{3/2}} \\ 
					\textrightarrow 
					71S\textsubscript{\scalebox{0.8}{1/2}} } 
				& \tabincell{c}{3.8 \textmu W /\\ 60.4 mW} 
				& \tabincell{c}{780 nm /\\ 480 nm} 
				& \tabincell{c}{100 MHz -\\ 1.4 GHz} 
				& \tabincell{l}{Detect over 1 GHz continuous-band with\\ best sensitivity of 1.5 \textmu V/cm/$\sqrt{\text{Hz}}$ and\\ 80 dB linear dynamic range.}\\

				\Xhline{0.6pt}
				\tabincell{c}{\cite{allinson2024simultaneous}} 
				& \tabincell{c}{CRS} 
				& \tabincell{c}{C1, C5\\ (multiband)} 
				& \tabincell{c}{F1, F2,\\ F6} 
				& \tabincell{c}{\textsuperscript{\scalebox{0.8}{133}}Cs} 
				& \tabincell{c}{
					6S\textsubscript{\scalebox{0.8}{1/2}} \textrightarrow 6P\textsubscript{\scalebox{0.8}{3/2}} \textrightarrow 
					19D\textsubscript{\scalebox{0.8}{5/2}} \textrightarrow \\
					17F\textsubscript{\scalebox{0.8}{7/2}} \textrightarrow 
					17G\textsubscript{\scalebox{0.8}{9/2}} \textrightarrow 17H\textsubscript{\scalebox{0.8}{11/2}} \\ \textrightarrow 17I\textsubscript{\scalebox{0.8}{13/2}} \textrightarrow 17K\textsubscript{\scalebox{0.8}{15/2}} \textrightarrow \\ 17L\textsubscript{\scalebox{0.8}{17/2}} \textrightarrow 17M\textsubscript{\scalebox{0.8}{19/2}} } 
				& \tabincell{c}{20 \textmu W /\\ 21 mW} 
				& \tabincell{c}{852 nm /\\ 519 nm} 
				& \tabincell{c}{0.607 THz,\\ 34.9 GHz,\\ 6.01 GHz, 1.85 GHz,\\ 700, 310, 128 MHz} 
				& \tabincell{l}{The CRS scheme was shown for simul-\\taneously detecting multiband signals.}\\
				
				\Xhline{0.6pt}
				\tabincell{c}{\cite{meyer2023simultaneous}} 
				& \tabincell{c}{DRFS,\\ superheterodyne} 
				& \tabincell{c}{C1, C2, C4\\ (QPSK), C5\\ (multiband)} 
				& \tabincell{c}{F1, F2,\\ F6} 
				& \tabincell{c}{\textsuperscript{\scalebox{0.8}{87}}Rb} 
				& \tabincell{c}{
					5S\textsubscript{\scalebox{0.8}{1/2}} \textrightarrow 5P\textsubscript{\scalebox{0.8}{3/2}} \textrightarrow \\
					56D\textsubscript{\scalebox{0.8}{5/2}} \textrightarrow 
					57P\textsubscript{\scalebox{0.8}{3/2}} \\ 
					(54F\textsubscript{\scalebox{0.8}{7/2}}, 59P\textsubscript{\scalebox{0.8}{3/2}}, 52F\textsubscript{\scalebox{0.8}{7/2}}) } 
				& \tabincell{c}{3.8 \textmu W /\\ 60.4 mW} 
				& \tabincell{c}{780 nm /\\ 480 nm} 
				& \tabincell{c}{1.72, 12.11 GHz,\\ 27.42, 65.11 GHz,\\ 115.75 GHz} 
				& \tabincell{l}{The DRFS scheme was shown for simul-\\taneously detecting five RF tones from\\ 1.7 GHz to 116 GHz.}\\

				\Xhline{0.6pt}
				\tabincell{c}{\cite{holloway2021multiple}} 
				& \tabincell{c}{MAS} 
				& \tabincell{c}{C1, C4 (AM, FM),\\ C5 (multiband)} 
				& \tabincell{c}{F1, F2,\\ F3, F6} 
				& \tabincell{c}{\textsuperscript{\scalebox{0.8}{133}}Cs\\ +\\ \textsuperscript{\scalebox{0.8}{85}}Rb} 
				& \tabincell{c}{
					(6S\textsubscript{\scalebox{0.8}{1/2}} \textrightarrow 6P\textsubscript{\scalebox{0.8}{3/2}} \textrightarrow \\
					34D\textsubscript{\scalebox{0.8}{5/2}} \textrightarrow 
					35P\textsubscript{\scalebox{0.8}{3/2}}), \\
					(5S\textsubscript{\scalebox{0.8}{1/2}} \textrightarrow 5P\textsubscript{\scalebox{0.8}{3/2}} \textrightarrow \\
					47D\textsubscript{\scalebox{0.8}{5/2}} \textrightarrow 
					48P\textsubscript{\scalebox{0.8}{3/2}}) } 
				& \tabincell{c}{(41.2 \textmu W /\\ 48.7 mW),\\
					(22.3 \textmu W /\\ 43.8 mW) } 
				& \tabincell{c}{(850.53 nm /\\ 511.148 nm),\\
					(780.24 nm /\\ 480.27 nm) } 
				& \tabincell{c}{19.626 GHz,\\
					20.644 GHz } 
				& \tabincell{l}{A multiband Rydberg atomic receiver is\\ achieved using the combination of two\\ different atomic species.}\\

			\Xhline{0.6pt}
			\rowcolor{yellow!60} 
			\multicolumn{10}{|c|}{\textbf{Wireless communication and sensing applications from the communication community}} \\

			\Xhline{0.6pt}
			\tabincell{c}{\cite{zhang2023quantum}} 
			& \tabincell{c}{Superheterodyne} 
			& \tabincell{c}{C1, C2, C6\\ (spatial\\ displacement)} 
			& \tabincell{c}{F1, F2} 
			& \tabincell{c}{\textsuperscript{\scalebox{0.8}{133}}Cs} 
			& \tabincell{c}{
				6S\textsubscript{\scalebox{0.8}{1/2}} \textrightarrow 6P\textsubscript{\scalebox{0.8}{3/2}} \textrightarrow \\
				66D\textsubscript{\scalebox{0.8}{5/2}} \textrightarrow 
				67P\textsubscript{\scalebox{0.8}{3/2}},\\
				(52D\textsubscript{\scalebox{0.8}{5/2}} \textrightarrow 
				53P\textsubscript{\scalebox{0.8}{3/2}}),\\
				(61D\textsubscript{\scalebox{0.8}{5/2}} \textrightarrow 
				63P\textsubscript{\scalebox{0.8}{3/2}}) } 
			& \tabincell{c}{N/A} 
			& \tabincell{c}{852 nm /\\ 510 nm} 
			& \tabincell{c}{2.4 GHz,\\ 5.0 GHz,\\ 28 GHz} 
			& \tabincell{l}{Improve sensing granularity by an order\\ of magnitude for WIFI signals and the\\ 28 GHz millimeter wave.}\\

			\Xhline{0.6pt}
			\tabincell{c}{\cite{gong2024rydberg_model}} 
			& \tabincell{c}{Superheterodyne} 
			& \tabincell{c}{C1, C2, C4} 
			& \tabincell{c}{F1, F2} 
			& \tabincell{c}{\textsuperscript{\scalebox{0.8}{133}}Cs} 
			& \tabincell{c}{
				6S\textsubscript{\scalebox{0.8}{1/2}} \textrightarrow 6P\textsubscript{\scalebox{0.8}{3/2}} \textrightarrow 
				47D\textsubscript{\scalebox{0.8}{5/2}} \\ \textrightarrow 
				48P\textsubscript{\scalebox{0.8}{3/2}} 
				(45D\textsubscript{\scalebox{0.8}{5/2}} \textrightarrow 46P\textsubscript{\scalebox{0.8}{3/2}}), \\
				(43D\textsubscript{\scalebox{0.8}{5/2}} \textrightarrow 44P\textsubscript{\scalebox{0.8}{3/2}}), 
				(40D\textsubscript{\scalebox{0.8}{5/2}} \\ \textrightarrow 41P\textsubscript{\scalebox{0.8}{3/2}}), 
				(66S\textsubscript{\scalebox{0.8}{1/2}} \textrightarrow 66P\textsubscript{\scalebox{0.8}{3/2}}), \\
				(63S\textsubscript{\scalebox{0.8}{1/2}} \textrightarrow 63P\textsubscript{\scalebox{0.8}{3/2}}) } 
			& \tabincell{c}{29.8 \textmu W\\ (9.4 \textmu W)\\ / 17 mW} 
			& \tabincell{c}{852 nm /\\ 510 nm} 
			& \tabincell{c}{6.9458, 7.9752,\\ 9.2186, 11.6187,\\
				13.4078, 15.5513 GHz} 
			& \tabincell{l}{Propose a RAQR-SISO scheme, signal\\ models, and provide performance analysis\\ for wireless communication and sensing.}\\
			
			\Xhline{0.6pt}
			\tabincell{c}{\cite{gong2024rydberg_DOA}} 
			& \tabincell{c}{Superheterodyne} 
			& \tabincell{c}{C1, C2, C6 (DOA)} 
			& \tabincell{c}{F1, F2} 
			& \tabincell{c}{\textsuperscript{\scalebox{0.8}{133}}Cs} 
			& \tabincell{c}{
				6S\textsubscript{\scalebox{0.8}{1/2}} \textrightarrow 6P\textsubscript{\scalebox{0.8}{3/2}} \textrightarrow \\
				47D\textsubscript{\scalebox{0.8}{5/2}} \textrightarrow 
				48P\textsubscript{\scalebox{0.8}{3/2}} } 
			& \tabincell{c}{29.8 \textmu W\\ / 17 mW} 
			& \tabincell{c}{852 nm /\\ 510 nm} 
			& \tabincell{c}{6.9458 GHz} 
			& \tabincell{l}{Propose a RAQ-MIMO wireless sensing\\ scheme, signal model, and a multi-target\\ DOA estimation algorithm.}\\

			\Xhline{0.6pt}
	\end{tabular}}
	
	\end{table*}

	\subsection{Verified Capabilities}
	
	\begin{itemize}

		\item 
		\textbf{C1: Amplitude Detection.}
		RAQRs are capable of measuring the amplitude of an RF signal using the standard scheme discussed in Section \ref{sec:standard}.
		
		\item 
		\textbf{C2: Phase Detection.}
		It can also retrieve the phase of an RF signal by harnessing the superheterodyne principle of Section \ref{sec:superheterodyne}.
		
		\item 
		\textbf{C3: Polarization Detection.}
		The polarization of an RF signal can be detected from the spectroscopic EIT signal by exploiting the PSS scheme of Section \ref{sec:polarization}.

		\item 
		\textbf{C4: Modulation Identification.}
		The amplitude and phase extraction capabilities allow RAQRs to identify RF signal waveforms relying on different modulations, such as analog modulation schemes: amplitude/frequency modulation (AM/FM) \cite{holloway2021multiple}, and digital modulation schemes: phase shift keying (PSK) (BPSK, QPSK \cite{holloway2019detecting, liu2022continuous}) as well as quadrature amplitude modulation (QAM) \cite{holloway2019detecting}.

		\item 
		\textbf{C5: Multiband/Continuous-Band Detection.}
        Simultaneously detecting multiband RF signals becomes feasible using a single RAQR. This can be realized by the CRS \cite{allinson2024simultaneous}, DRFS \cite{meyer2023simultaneous}, and MAS \cite{holloway2021multiple} schemes of Section \ref{sec:MDBD}. The multiple bands can span over a continuous frequency range through the ARRT scheme of Section \ref{sec:CBD}.

		\item 
		\textbf{C6: Spatial Displacement and Direction Detection.}
		The spatial displacement and direction-of-arrival (DOA) of a moving target can be detected by RAQRs. By recording the RF signal phase-change caused by target movements, one can deduce spatial displacement \cite{zhang2023quantum}. By acquiring the phase-differences of the RF signal between different locations, DOA estimation can be realized \cite{gong2024rydberg_DOA}.

	\end{itemize}

	\subsection{Distinctive Features}

	\begin{itemize}
		\item 
		\textbf{F1: Extremely-High Sensitivity.} 
        Compared to traditional RF receivers, RAQRs exhibit ultra-high sensitivity in detecting RF signals, as discussed in Section \ref{sec:standard}, \ref{sec:superheterodyne}, and \ref{sec:OLA}. The sensitivity can be on the order of $<$ nV/cm/$\sqrt{\text{Hz}}$ \cite{zhang2024rydberg,wu2024enhancing}, surpassing the 1.5 nV/cm/$\sqrt{\text{Hz}}$ sensitivity of metal antennas \cite{zhang2024rydberg}.

		\item 
		\textbf{F2: Broadband Tunability.} 
        RAQRs allow a multitude of options in choosing different atomic energy levels combined with different electron transitions to realize the detection of RF signals across a broad frequency range (DC to THz) using a single receiver \cite{Fancher2021Rydberg}.

		\item 
		\textbf{F3: Narrowband Selectivity.}
        For a standard RAQR scheme, the atomic response has a narrowband selectivity ($\le$ 10 MHz \cite{Fancher2021Rydberg}), where only a narrow instantaneous bandwidth around the carrier frequency can be realized for coupling between adjacent Rydberg states \cite{schlossberger2024rydberg,Fancher2021Rydberg}.

		\item 
		\textbf{F4: Enhanced Instantaneous Bandwidth.}
		The instantaneous bandwidth can be significantly extended by exploiting the ARRT scheme \cite{simons2021continuous}, which can be up to $100$ MHz \cite{zhang2024rydberg} and reference [39] of \cite{zhang2024rydberg}, as discussed in \ref{sec:CBD}. This facilitates RAQRs in receiving RF signals having a large continuous bandwidth.

		\item 
		\textbf{F5: Ultra-Wide Input Power Range.}
        RAQRs are capable of receiving not only very weak signals, but also very strong signals up to several kV/m \cite{Fancher2021Rydberg}. This ensures the receiver's operation in high-intensity RF environments.

		\item 
		\textbf{F6: Ultra-High Scalability.}
        RAQRs exhibit an ultra-high scalability to form advanced receivers (e.g., arrays, multiband, continuous-band, and PSS receivers) using a single vapor cell based on the schemes discussed in Section \ref{Sec:Principal&Scheme}. These functionalities can be potentially integrated with a single vapor cell to realize multi-functional receivers.

	\end{itemize}

 \begin{figure*}[t!]
 	\centering
 	\includegraphics[width=0.92\textwidth]{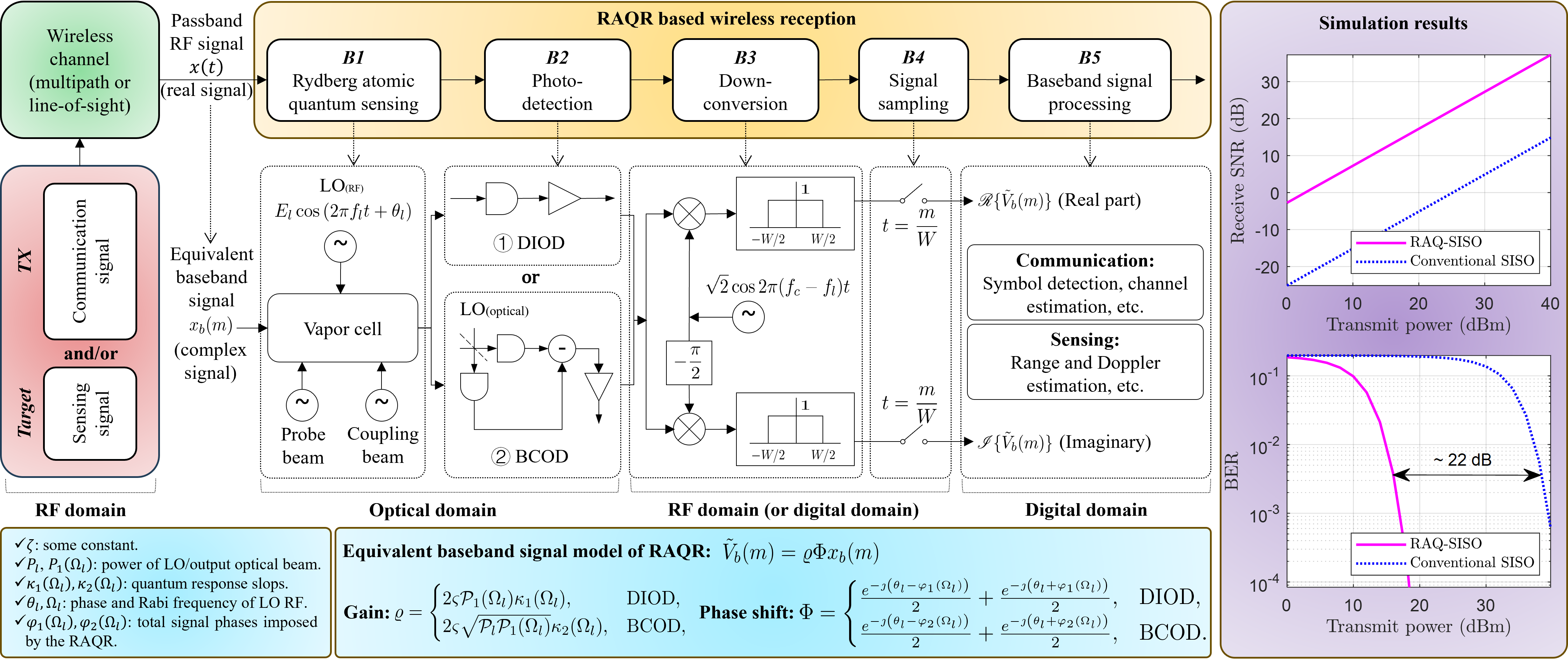}
 	\vspace{-0.5em}
 	\caption{Superheterodyne RAQR based wireless reception scheme, signal model, and simulation results.}
 	\label{fig:TRStructure}
 	\vspace{-1em}
 \end{figure*}

	\subsection{State-of-the-Art in RAQRs}
	Next, we review several representative works from both the physics and communication communities (seen in TABLE \ref{tab:State-of-the-art}).

	\subsubsection{Theoretical and Experimental Studies From the Physics Community}
	The studies on RAQRs from the physics community can be classified into two branches. The first branch is from a theoretical perspective, aiming to analyse the fundamental performance limit of SQL and practical impairments on RAQRs \cite{schlossberger2024rydberg,Fancher2021Rydberg}, and to construct specific quantum response models for novel RAQR concepts, such as many-body Rydberg atom \cite{zhang2024rydberg}. 
	Another branch is experimentally designing novel atomic structures to support new capabilities, demonstrate specific features, explore new sensitivity limits, and so forth \cite{jing2020atomic, wu2024enhancing,elgee2024complete,simons2021continuous,liu2022continuous,allinson2024simultaneous,meyer2023simultaneous,holloway2021multiple, holloway2019detecting}. The advances of RAQRs have confirmed the feasibility of realizing powerful RF receivers, which paves the way for more complex upper-layer communication and sensing applications.

	\subsubsection{Wireless Communication and Sensing Applications From the Communication Community}
	Inspired by recent advances of RAQRs, wireless communication and sensing applications started to emerge. Reference \cite{zhang2023quantum} experimentally demonstrated RAQRs' superiority in sensing moving objects. Then, the work \cite{gong2024rydberg_model} took the lead in constructing useful signal models and analysing the superior performance of RAQR-aided wireless systems. Furthermore, the work \cite{gong2024rydberg_DOA} exploited a Rydberg receiver array for realizing multi-target DOA estimation by proposing a corresponding DOA estimation algorithm.

    \section{RAQ-SISO and RAQ-MIMO Schemes}
	\label{Section_PCM}
	
	In this section, we introduce the RAQRs into classical wireless systems by detailing the receiver structure of RAQ-SISO systems, and presenting potential RAQ-MIMO schemes in both centralized and distributed applications.

	\subsection{Reception Scheme of RAQ-SISO Systems}
        \label{subsec:TRA}
	Fig. \ref{fig:TRStructure} shows each stage of the holistic RAQ-SISO system and its corresponding signal model from an equivalent baseband communication perspective. The received input signal of the RAQR may be a communication signal from a classical wireless transmitter or a sensing signal reflected from some moving target. The received signal is assumed to have a carrier frequency of $f_c$, which is processed by the following blocks.

    \begin{enumerate}[label={\em \textbf{B\arabic*}}, leftmargin=1.7em, labelindent=0pt, itemindent = 0em]
		\item \label{itm:1} 
		\textbf{Rydberg atomic quantum sensing:} It realizes RF reception using superheterodyne RAQRs. Its optical output is modulated and can be approximated by an amplified cosine wave of frequency $f_c - f_l$ \cite{jing2020atomic,gong2024rydberg_model}. 
		\item \label{itm:2} 
		\textbf{Photodetection:} The optical signal is detected by a PD, which can be realized by direct incoherent optical detection (DIOD) or balanced coherent optical detection (BCOD) schemes \cite{gong2024rydberg_model}.
        Both schemes generate an output that is approximated by a cosine wave of frequency $f_c - f_l$.
        \item \label{itm:2} 
		\textbf{Down-conversion:} To obtain a continuous-time baseband signal of the PD's output, down-conversion is realized by a homodyne receiver (HR), which requires IQ mixers having a mixing frequency of $f_c - f_l$ and lowpass filters having a bandwidth of $W$ (nearly the same as the signal bandwidth). We emphasize that this functionality can be implemented in either RF or digital domains. 
        \item \label{itm:2} 
		\textbf{Signal sampling:} The continuous-time baseband signal is sampled by analog-to-digital converters (ADCs) and then the complex baseband samples are obtained. The sampling rate obeys Nyquist's sampling theorem.
        \item \label{itm:2} 
		\textbf{Baseband signal processing:} The baseband samples embody both the intensity and phase of the desired RF signal, which are utilized for recovering the transmitted baseband signal or for estimating the information of the target using specific signal processing algorithms.
    \end{enumerate}

    The equivalent baseband signal model of the holistic wireless system constructed in \cite{gong2024rydberg_model} is briefly summarized at the bottom of Fig. \ref{fig:TRStructure}, where the parameters are explained in the box at the bottom. It is seen that the RAQR imposes a gain $\varrho$ and a phase shift $\Phi$ on the received RF signal. These parameters are determined by both the quantum response and the photodetection scheme selected. The accuracy of this signal model has been verified in \cite{gong2024rydberg_model}. 
    In our simulations of this article, we follow the same parameter configurations of the unoptimized RAQR and of the conventional RF receiver as in \cite{gong2024rydberg_model} and showcase the received SNR and bit-error-rate (BER) versus the transmit power under Rayleigh fading channels. The transmission distance is 200 meters having a path loss factor of 3.8. The results reveal that the RAQ-SISO system significantly outperforms the conventional SISO systems under the same transmit power, as seen in the simulation results of Fig. \ref{fig:TRStructure}.

    \begin{figure*}[t!]
    	\centering
    	\includegraphics[width=0.95\textwidth]{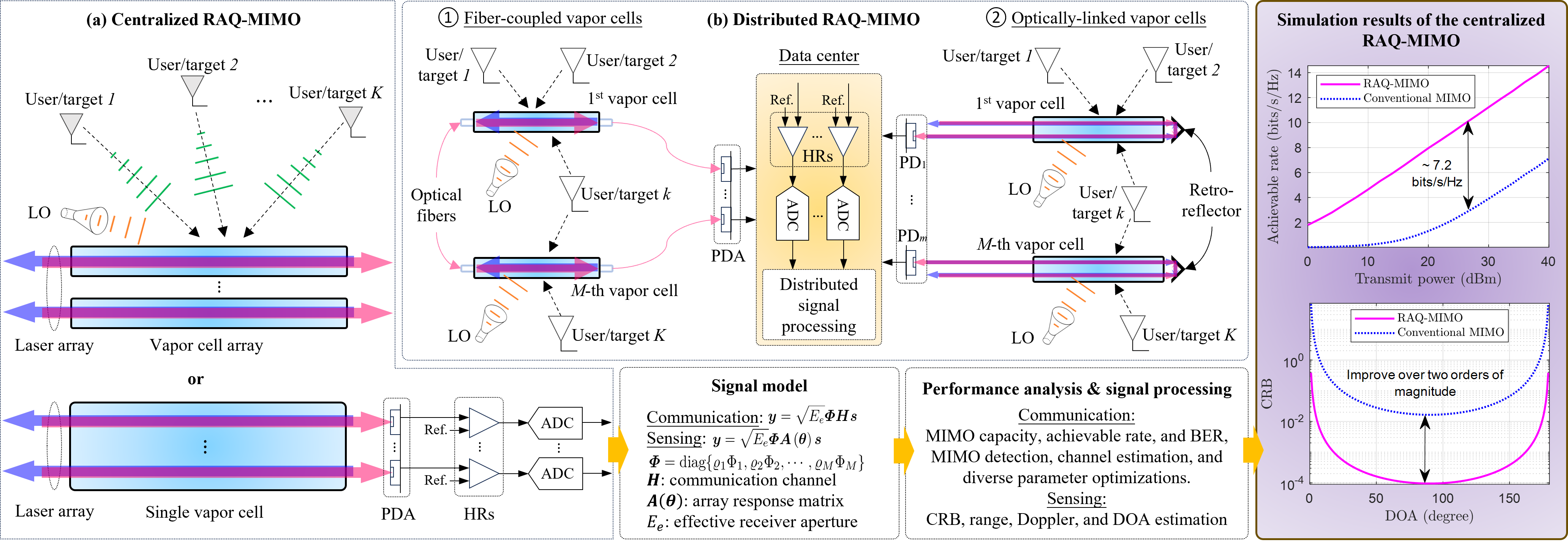}
    	\caption{(a) Centralized RAQ-MIMO, (b) distributed RAQ-MIMO, and (c) simulation results of the centralized RAQ-MIMO.}
    	\label{fig:AtomicQuantumMIMO}
    	\vspace{-1em}
    \end{figure*}

	\subsection{Centralized and Distributed RAQ-MIMO Schemes}
	
	\subsubsection{Centralized RAQ-MIMO}
	The most straightforward way to form a co-located array is using multiple vapor cells, where each vapor cell is penetrated using a probe-coupling beam pair to form the receive region, as seen at the top of Fig. \ref{fig:AtomicQuantumMIMO}(a). Furthermore, we introduce a more compact one shown at the bottom of this figure, which is implemented by penetrating a single vapor cell using a laser beam array. The former scheme has the advantage of constructing large receiver apertures, while the latter scheme facilitates compact receiver apertures.  For both schemes, their parallel optical outputs are forwarded to the corresponding photodetector array (PDA), HRs, and ADCs. Therefore, all samples can be jointly modelled to obey a corresponding signal model. These samples are then appropriately combined to realize the desired communication/sensing tasks by using specific algorithms, as seen in the `Signal model' and `Performance analysis \& signal processing' boxes of Fig. \ref{fig:AtomicQuantumMIMO}(a). 
	
	For both of the above schemes, the optical array signals experience diversity gains owing to their different locations. Compared to the OLA scheme \cite{wu2024enhancing}, these centralized RAQ-MIMO schemes combine all branch signals in the digital domain instead of directly combining all the output optical signals. This retains higher degrees-of-freedom for the information of signal and facilitates better performance.  Furthermore, we present simulation results of a single vapor cell having a uniform linear array of 5 laser beams (5 receive regions), where each region is identically configured as that of the RAQ-SISO. The achievable rate of wireless communication and the Cramér–Rao bound (CRB) of DOA estimation are showcased for a single user/target at a distance of 200 meters. As observed from the simulation results of Fig. \ref{fig:AtomicQuantumMIMO}(a), the achievable rate of RAQ-MIMO is $\sim 7.2$ bits/s/Hz higher than that of the conventional MIMO systems under Rayleigh fading channels. Additionally, the CRB achieved by the RAR-MIMO improves that achieved by the conventional MIMO by as much as two orders of magnitude.

	\subsubsection{Distributed RAQ-MIMO}
	To further enhance the spatial diversity and improve the coverage, we present the distributed RAQ-MIMO scheme, where the system consists of multiple vapor cells and a single data center. Specifically, the vapor cells are spatially distributed across a long distance range, so that each vapor cell is capable of receiving RF signals having larger spatial diversity or covering more users/targets. The data center consists of laser generators, the PDA/multiple PDs (responsible for receiving long-distant optical signal), the HRs, ADCs, and the distributed signal processing unit. 
	
	The data center and vapor cells are connected in two ways, namely the fiber-coupled and optically-linked schemes of Fig. \ref{fig:AtomicQuantumMIMO}(b). These schemes are based on the recent advances of fiber-coupled vapor cells \cite{schlossberger2024rydberg} (references [117], [119] therein) and of passive RAQRs at long distances \cite{otto2023distant}. The two schemes have their own advantages and shortcomings. Specifically, the fiber-coupled scheme appear to be more robust to environment uncertainty than the opticall-linked scheme due to the higher reliability of fibers, while the latter exhibits higher mobility and is easier to be deployed.

	\section{Open Problems and Future Directions}
	\label{Section_FRD}

    Advanced designs of RAQRs have the potential of approaching or even exceeding the SQL, thereby outperforming the Shannon-limit of the current RF receivers. 
    Even though significant advances have been achieved in the evolution of RAQRs, the answers to several open problems remain to be further explored. Therefore, we list a number of future directions of significant interest as follows.

		\textbf{Improvements of Receiver Metrics and Commercialization of RAQRs:} Albeit the basic capabilities of RAQRs have been explored and verified. Future studies are required to be continued for pushing forward the receiver performance, such as the sensitivity, bandwidth, and robustness, etc. Furthermore, to commercialize RAQRs, mature cost-saving, miniaturization, and power-conserving technologies are required.

		\textbf{Applications of RAQRs to Next-Generation Wireless Communication and Sensing:} To fully harness the benefits of RAQRs, their integration into wireless communication and sensing is high of significance. At the time of writing, the integration is in its infancy. More particularly, we expect three main directions to be explored. 
		\begin{enumerate}[label={\em \textbf{D\arabic*}}, leftmargin=1.7em, labelindent=0pt, itemindent = 0em]
			\item \label{itm:d1} 
			\textbf{Theoretical Modelling of RAQR-Aided Wireless Systems:} 
			The currently available theoretical modelling is mainly dedicated to the input-output relationship of the vapor cell. However, the holistic baseband equivalent models have to be developed for various wireless systems (e.g., wideband, MIMO).

			\item \label{itm:d2} 
			\textbf{Signal Processing of RAQR-Aided Wireless Systems:} 
			With the aid of mathematical models constructed, computationally efficient signal processing algorithms must be conceived for communication-/sensing-oriented functionalities, including signal detection, channel estimation, and spatial displacement/direction detection etc.

			\item \label{itm:d3} 
			\textbf{Interplay between RAQRs and Classical Transmitters:} 
			In RAQR-aided wireless systems, classical antenna-based transmitters are still needed. The interesting interplay between RAQRs and classical transmitters opens a new research direction on the design of novel transceivers.

			\item \label{itm:d4} 
			\textbf{System-Level Experimental Implementations and Validations:}
			Previous efforts have shown the great potential of RAQRs, but substantive future efforts are required for experimentally investigating their practical system-level capabilities and benefits.
			
		\end{enumerate}

	\section{Conclusions}
	\label{Section_CON}
	In this article, RAQRs and their integration into wireless communication and sensing have been considered. We have briefly introduced the fundamentals of Rydberg atoms and discussed the RAQRs' capabilities as well as distinctive features, followed by a contemporary survey of the state-of-the-art studies. To support the integration of RAQRs with classical wireless systems, we have advocated the RAQ-SISO and RAQ-MIMO schemes. Finally, we have concluded by proposing promising research directions.


	\bibliographystyle{IEEEtran}
	\bibliography{IEEEabrv,references} 

\end{document}